*Sequence Analysis*

# EDGE COVID-19: A Web Platform to generate submission-ready genomes from SARS-CoV-2 sequencing efforts

Chien-Chi Lo[1]*, Migun Shakya[1], Karen Davenport, Mark Flynn, Adán Myers y Gutiérrez, Bin Hu, Po-E Li, Elais Player Jackson, Yan Xu, and Patrick S. G. Chain*

Bioscience Division, Los Alamos National Laboratory, Los Alamos, New Mexico

[1]Contributed equally to this work.

*To whom correspondence should be addressed.

## Abstract

**Summary:** Genomics has become an essential technology for surveilling emerging infectious disease outbreaks. A wide range of technologies and strategies for pathogen genome enrichment and sequencing are being used by laboratories worldwide, together with different, and sometimes ad hoc, analytical procedures for generating genome sequences. A standardized analytical process for consensus genome sequence determination, particularly for outbreaks such as the ongoing COVID-19 pandemic, is critical to provide a solid genomic basis for epidemiological analyses and well-informed decision making. We have developed a bioinformatic workflow to standardize the analysis of SARS-CoV-2 sequencing data generated with either the Illumina or Oxford Nanopore platforms. Using an intuitive web-based interface, this workflow automates SARS-CoV-2 reference-based genome assembly, variant calling, lineage determination, and provides the ability to submit the consensus sequence and necessary metadata to GenBank, GISAID, and INSDC raw data repositories.
**Availability:** https://edge-covid19.edgebioinformatics.org, and https://github.com/LANL-Bioinformatics/EDGE/tree/SARS-CoV2
**Contact:** Chien-Chi Lo (chienchi@lanl.gov) and Patrick Chain (pchain@lanl.gov)
**Supplementary information:** Supplementary data are available at *Bioinformatics* online.

## 1 Introduction

Public health laboratories and scientists around the world have been sequencing the genome of SARS-CoV-2, the etiological agent responsible for the COVID-19 pandemic. In just over a year, more than 1 million genomes have been sequenced and made publicly available via genome repositories, such as GISAID (Shu and McCauley, 2017) and GenBank (Clark, et al., 2016). Multiple sequencing platforms (Illumina, Oxford Nanopore, and PacBio) and experimental approaches are being used to obtain these genomes, including enriching for the virus before sequencing, amplifying and sequencing overlapping genomic regions, or performing deep, random, 'shotgun' sequencing. While multiple strategies and platforms provide different trade-offs that can be tailored to address specific scientific questions, they complicate the development of a bioinformatic workflow that can process diverse input data to produce high quality genome sequences.

There are many available software options for each individual bioinformatic step in genome consensus generation, and often for each sequencing technology and/or approach. However, most of these can only be executed from the command-line in a Linux environment and are not readily available as an automated workflow. This accessibility barrier and lack of standardization (e.g., on consensus base calling and thresholds for underlying read data), presents a major challenge to most bench scientists without extensive training in bioinformatics, and can result in *ad hoc* procedures for deciding the final genome. Finally, the lack of automated genome sequence submission to public repositories results in paucity or delays in publicly available data.

There is a need for a standardized, easy-to-use, GUI-based, bioinformatics application that help automate the routine, though complex, bioinformatics steps required to generate high quality genomes for submission to public repositories. Such an application would position labs with limited computational resources and bioinformatics expertise to perform their own analyses, while encouraging standardization of SARS-CoV-2 genome processing. To address this need, we developed EDGE COVID-19 (EC-19), a user-friendly, web-based platform that enables rapid, automated, and standardized processing of FASTQ read datasets from either Illumina or Oxford Nanopore (ONT) platforms generated using either amplicon or random shotgun-based methods. The output of this workflow is a consensus genome, an inventory of all SNPs and variants, the predicted lineage of the consensus genome, and metadata information sufficient for submitting genomes to GenBank, GISAID, and Sequence Read Archive (SRA). This platform is freely available as a Docker container for local installation and as a public web-service, where users can register for accounts, upload data, run analyses, and download results.

## 2 Design, Implementation, and Features

EC-19 is a tailored version of the more generic open-source, EDGE bioinformatics platform (Li, et al., 2017) that was developed to democratize genomic analysis and make complex bioinformatic tools available to non-experts. Although it includes portions of the original platform, EC-19 has been customized for SARS-CoV-2 genomes.

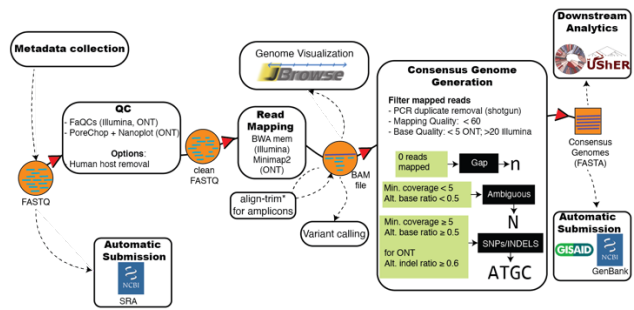

**Figure 1: Overview of EDGE COVID-19 workflow.** The workflow includes Quality Control, mapping reads to a SARS-CoV-2 reference genome sequence, removing primer sequences when needed, generation of consensus genomes, variant calling, lineage call, and phylogenetic placement. Dotted line indicates optional steps. More details that include underlying tools and their versions can be found in online documentation.

EC-19 allows users to input their own Illumina or ONT sequencing FASTQ files, or automatically obtained them from the INSDC (International Nucleotide Sequence Database Collaboration) raw read repositories. The workflow (**Figure 1**) includes quality control, mapping reads to a SARS-CoV-2 reference genome, removal of primer sequences, generation of consensus genomes, variant calling, and the ability to take sample metadata and automatically deposit genomes to GISAID and GenBank, and raw reads to INSDC. A slightly modified SARS-CoV-2 RefSeq genome, NC_045512.2 (33 poly-A tail from the 3' end of the genome is removed) is used as the default reference. Users can also specify additional SARS-CoV-2 genome from GenBank, or upload their own reference to generate a consensus genome. Additionally, to account for higher error rate in ONT data, particularly within homopolymer nucleotides, we use a more stringent ONT specific 60% threshold for detection of insertions/deletions (indels) (Cretu Stancu, et al., 2017; Rang, et al., 2018). Detailed parameters for each step can be found in the online documentation, most of which are able to be modified by the users as well.

Outputs of the workflow include statistics and figures for quality control, read mapping, and consensus genome generation, as well as all original output files from any tool used during the analysis. The integrated genome browser JBrowse (Buels, et al., 2016) also enables a deeper, visual inspection of the underlying data.

## 3 Genome Processing

Dozens of SARS-CoV-2 genomes sequenced as part of a surveillance program in New Mexico, USA have been successfully processed and submitted to both GISAID and GenBank genome repositories using EC-19 (**Supplementary Table 1**). Individual samples were sequenced using either ONT or Illumina platforms with either the ARTIC or SWIFT PCR enrichment protocols. The EC-19 results can also be accessed from the public project list on the EC-19 website (https://edge-covid19.edgebioinformatics.org/).

## 4 Conclusion

While the SARS-CoV-2 pandemic has both highlighted the utility of genome sequencing and our collective ability to generate enormous numbers of genomes during an outbreak, it is often unclear how individual genome sequences have been generated and how much confidence to assign to consensus sequences, in part or in whole. This is rendered enormously complex by the number of different sequencing protocols available, the number of different sequencing technologies used, the number of different bioinformatics strategies being applied, and the number of different labs (with diverse expertise and experience with genomics) generating the data. Efforts such as the EC-19 web platform are meant to lower the barrier for submitting genomes for public use, while helping to standardize and generate high quality genomes. EC-19 provides a potential solution to democratizing genomic surveillance capabilities for the global community by using a robust suite of standardized bioinformatics workflows in a user-friendly web browser.

## Funding

This work was supported by the Los Alamos National Laboratory LDRD (20200732ER), DTRA (CB10152 and CB10623), and DOE (KP160101 and 4000150817). Hosting of edgebioinformatics.org is provided by Cyverse, which is supported by the NSF (DBI-0735191, DBI-1265383, and DBI-1743442).

*Conflict of Interest:* none declared.